\documentclass[
preprint,
showpacs,
showkeys
]{revtex4-1}
\usepackage{amsthm}
\theoremstyle{plain}
\newtheorem*{theorem*}{Theorem}

\newtheorem*{definition*}{Definition}

\newtheorem*{lemma*}{Lemma}

\usepackage{braket}
\usepackage{delarray}
\usepackage{here}

\usepackage{txfonts}

\usepackage[dvipdfmx]{graphicx}
\usepackage{bm}
\usepackage{color}
\usepackage{ulem}

\newcommand{\be}{\begin{eqnarray}}
\newcommand{\ee}{\end{eqnarray}}
\newcommand{\ba}{\begin{array}}
\newcommand{\ea}{\end{array}}
\newcommand{\bmat}{\left(\begin{array}}
\newcommand{\emat}{\end{array}\right)}

\begin{document}
\title{
Control limit for the quantum state preparation under stochastic control errors
}

\author{Kohei Kobayashi$^1$}

\affiliation{$^1$Global Research Center for  Quantum Information Science, National Institute of Informatics,
 2-1-2 Hitotsubashi,  Chiyoda-ku, Tokyo 101-8340, Japan}

\begin{abstract} 
We investigate the effect of stochastic control errors on the Hamiltonian that controls a closed quantum system.
Quantum information technologies require careful control for preparing a desired  state used as an information resource.
However, because the stochastic control errors inevitably appear in realistic situation,
 it is difficult to completely implement the control Hamiltonian.
Under this error, the actual performance of quantum control is far away from the ideal one, and thus
 it is of great importance to evaluate the effect of the control errors.
In this paper, we derive a lower bound of the fidelity between two closed quantum systems obeying the dynamics with and without errors.
This bound reveals a reachable and unreachable set of the controlled quantum system under stochastic noises.
Also, it is easily computable without considering the stochastic process and needing the full dynamics of the states.
We demonstrate the actual performance of this bound via a simple control example. 
Furthermore, based on this result, we quantitatively evaluate the probability of obtaining the target state in the presence of control errors.


\end{abstract}
\date{\today}
\maketitle

\section{Introduction}

Recently, there has been considerable interests in quantum information technologies including quantum computing. 
To realize these technologies, it is necessary to prepare a desired quantum state used as an information resource.
Thus, an accurate technique for manipulating the quantum system and preparing the desired target state plays an essential role \cite{Nielsen}.
When the effect of external environments such as decoherence can be ignored,
the ideal time evolution of the quantum system is described by the Schr${\rm \ddot{o}}$dinger equation. 
Therefore, the target state can be prepared by suitably implementing the control Hamiltonian.
For example, the scheme of quantum annealing \cite{qa1, qa2, qa3, qa4} or quantum adiabatic computation \cite{qc1, qc2, qc3, qc4, qc5} 
are modeled by the Schr${\rm \ddot{o}}$dinger equation.

However, it is difficult to completely implement the control Hamiltonian without any errors in realistic situation.
In particular, considering experimental control setup, analog control errors on the parameters are inevitable.
Mainly, there are two types of analog control errors;
deterministic control error and stochastic control error.
Deterministic control error that deterministically effect on quantum dynamics, such as a bias of the mafnetic field, 
have been investigated so far in several literatures \cite{dn1, dn2, dn3}.
Also, stochastic control errors, such as a stochastic fluctuation occurring on the control Hamiltonian at each time,
is one of the critical obstacles to overcome \cite{sn1, sn2, sn3, sn4, sn5}.
In unitary dynamics, stochastic errors can be formulated as a time-varying stochastic noise, and 
the time evolution under it can be described by the stochastic differential equation.
Although some researhes have studies the stochastic noise \cite{sn3, sn4, sn5},
 no studies have investigated its effect in a general and rigorous manner.

This paper aims to quantively investigate the influence of the stochastic errors on quantum state preparation.
We present a lower bound of the fidelity
between the two closed quantum states obeying the dynamics with and without stochastic control errors. 
This bound is computable without needing the full dynamics of the states.
Also it gives a quantitative limit on the Hamiltonian control for preparing the quantum state under control errors as a function of the noise strength.
Furthermore, based on this result, we quantitatively evaluate the probability of obtaining the target state in the presence of control errors.

\section{Lower bound for the closed system under stochastic noise}

\subsection{Dynamics of the closed system}

We begin with the explanation of the setup.
Let $\ket{\psi(t)}$ be the quantum state in the absence of the stochastic control errors, which obeys the following equation:
\begin{equation}
\label{SE}
\frac{d \ket{\psi(t) }}{dt}=-i \hat{H} (t)\ket{\psi(t)}, \ \ \ \ket{\psi(0)}=\ket{\psi_0},
\end{equation}
where $\hat{H}(t)$ is the time-dependent Hamiltonian that controls the quantum system (we set $\hbar=1$). 
We assume that $\ket{\psi(t)}$ reaches the target state $\ket{\psi(T)}$ at final time $t=T$.

Next, we consider the another state $\ket{\phi(t)}$ driven by $\hat{H}(t)$ and $\sum_k\hat{B}_k(t)$.
$\hat{B}_k(t)$ denotes a $k$th stochastic control error and satisfies $\hat{B}_k^\dagger(t)=\hat{B}_k(t)$.
In this case, it is known that the time evolution of $\ket{\phi(t)}$ is described by the following stochastic differential equation \cite{sn4}:

\begin{equation}
\label{Stratonovich}
d\ket{\phi(t)} =-\left(i\hat{H}(t)dt + \sum_k \hat{B}_k(t) \circ dW_k(t) \right)\ket{\phi(t) },
\end{equation}
where the symbol $\circ$ denotes the Stratonovich interpretation and $W_k(t)$ is the Gaussian Wiener process  satisfying $dW^2_k(t)=dt$.
The another expression of this stochastic process is written by 

\begin{equation}
\label{SSE}
d\ket{\phi(t)} =-\left(i\hat{H}(t) +\frac{1}{2} \sum_k \hat{B}^2_k(t) \right)\ket{\phi(t) }dt 
-i \sum_k\hat{B}_k(t) \ket{\phi(t) } \bullet dW_k(t),
\end{equation}
where the symbol $\bullet$ is the It\={o}  interpretation.
By modifying the stochastic process from the  Stratonovich form to the It\={o}  form, 
we can take the expectation $\mathbb{E}$, as will be seen later.
Moreover, we impose the following condition on $\hat{B}_k(t)$:
\begin{equation}
\label{errorcondition}
\hat{B}^2_k(t)=\gamma^2_k(t)\hat{I},
\end{equation}
where $\hat{I}$ denotes the identity operator and $\gamma_k(t)$ is the time-dependent noise strength.
For example, when $\hat{B}_k(t)$ is constructed from the tensor product of the Pauli matrices
$\hat{B}_k(t)=\gamma_k(t)\hat{S}\otimes \hat{S}\otimes \cdots$,
this error condition (\ref{errorcondition}) is satisfied.
However, for example, when $\hat{B}_k(t)$ corresponds to the spin angular momentum operator
$\hat{B}_k(t)=\gamma_k(t)\sum^N_j \hat{S}^{(j)}$ or the annihilation or creation operator,
Eq. (\ref{errorcondition})  is not satisfied.

Under the condition (\ref{errorcondition}), Eq. (\ref{SSE}) becomes
\begin{equation}
\label{noisydynamics}
d\ket{\phi(t)} =-\left(i\hat{H}(t) +\frac{1}{2} \sum_k \gamma^2_k(t) \hat{I}   \right)\ket{\phi(t)}dt -i\sum_k \hat{B}_k(t)\ket{\phi(t)} \bullet dW_k(t).
\end{equation}

Taking the expectation of the both sides of Eq. (\ref{noisydynamics}) with respect to $dW_k(t)$, due to $\mathbb{E}[dW_k(t)]=0$, we have 
\begin{equation}
\frac{d \mathbb{E}[ \ket{\phi(t)} ] }{dt}=-\left(i \hat{H} (t)+ \frac{1}{2} \sum_k \gamma^2_k(t) \hat{I} \right)\mathbb{E}[ \ket{\phi(t)}].
\end{equation}

Now we assume that the initial states of $\ket{\psi(t)}$ and $\ket{\phi(t)}$ are identical $\ket{\psi(0)}=\ket{\phi(0)}=\ket{\psi_0}$. 
Then, the desired final state $\ket{\psi(T)}$ is given by
 \begin{equation}
\ket{\psi(T)}={\rm exp}_+\left(-i\int^T_0 \hat{H}(t)dt \right)\ket{\psi_0},
\end{equation}
where ${\rm exp}_+$ denotes the time-ordered exponential.
From this $\ket{\psi(T)}$ and the fact that the terms $\gamma^2(t)\hat{I}$ commute with $\hat{H}(t)$ for all $t$,  
we can write the expectation state $\mathbb{E}[ \ket{\phi(T)} ]$ as follows:
 \begin{equation}
\mathbb{E}[ \ket{\phi(T)} ]= {\rm exp}\left( -\frac{1}{2}\int^T_0 \sum_k \gamma^2_k(t) dt\right) \ket{\psi(T)}.
\end{equation}
This indicates that the stochastic control error makes the final obtained state $\mathbb{E}[ \ket{\phi(T)} ]$ away from the ideal one exponentially.
Even if  $\sum_k\gamma_k(t)$ is small, as the driving time becomes longer, the control achievement degrades.
Therefore, it is important problem to quantify how much the preparation of a desired state can be achieved under stochastic noises.

\subsection{Derivation of the lower bound}

Consider the fidelity between $\ket{\psi(t)}$ and $\ket{\phi(t)}$:
 \begin{equation}
\label{fidelity}
F(t) :=|\langle \psi(t)|\phi(t) \rangle|^2,  
 \end{equation}
where $0\leq F(t)\leq 1$.
$F(t)$ is employed as a cost function in various quantum science scenario, 
because it can be easily computed both analytically and numerically \cite{Nielsen}. 
If and only if  $\ket{\psi(t)} =e^{i\theta}\ket{\phi(t) }$ ($\theta \in\mathbb{R}$ is a global phase factor), $F(t)=1$ is achieved.
Because $F(t)$ decreases from $1$ for $t>0$ under the control errors,
our aimed result is a lower bound of $F(t)$ such that $F(t)\geq F_*\geq0$.

We find the time evolution of the overlap  $\langle \psi(t) | \phi(t) \rangle$:
\begin{eqnarray}
\label{deriveupperbound2} 
d\left( \langle \psi(t) | \phi(t) \rangle \right) &=&\langle \psi(t) |i \hat{H}(t) | \phi(t) \rangle dt \nonumber \\
&&\ \ \ -\langle \psi(t) |\left(i \hat{H}(t)dt + \sum_{k}\frac{\gamma^2_k(t) }{2} \hat{I} dt +\sum_k i \hat{B}_k(t) \bullet dW_k(t)  \right)|\phi(t) \rangle  \nonumber \\
&=& -\sum_k \frac{\gamma^2_k(t) }{2}\langle \psi(t) | \phi(t) \rangle dt-\sum_k \langle \psi(t) |i \hat{B}_k(t) | \phi(t) \rangle\bullet dW_k(t).
\end{eqnarray}
Taking the real part and the ensemble average of this equation,
\begin{equation}
\frac{ d\mathbb{E}[ \Re \left\{ \langle \psi(t) | \phi(t)  )\rangle  \right\}] }{dt}
=-\sum_k \frac{\gamma^2_k(t) }{2} \mathbb{E}[ \Re \left\{ \langle \psi(t)| \phi(t) \rangle \right\} ].
\end{equation}
Solving this equation yields
\begin{eqnarray}
\label{deriveupperbound3} 
\mathbb{E}[ \Re\left\{ \langle \psi(t)| \phi(t) \rangle \right\} ]  ={\rm exp}\left( -\frac{1}{2}\int^t_0 \sum_k\gamma^2_k(t')dt'\right).
\end{eqnarray}
Further, by using the inequality $\mathbb{E}[x^2] \geq \mathbb{E}[x]^2$,
\begin{eqnarray}
\label{lowerbound}
\mathbb{E}[ \Re\left\{ \langle \psi(t) | \phi(t) \rangle \right\} ] &\leq&  \mathbb{E}[  |\langle \psi(t) | \phi(t) \rangle| ] \nonumber \\
&\leq& \sqrt{  \mathbb{E}[ |\langle \psi(t) | \phi(t)\rangle|^2 ]} \nonumber \\
&=&  \sqrt{ \mathbb{E}[ F(t) ] }.
\end{eqnarray}

From (\ref{deriveupperbound3})  and (\ref{lowerbound}), we end up with the following lower bound:

\begin{equation}
\label{theorem}
\mathbb{E}[ F(T) ]  \geq F_*:={\rm exp}\left(-\int^T_0 \sum_k \gamma^2_k(t)dt\right),
\end{equation}
for  $T\in[0, \infty)$. This inequality is the main result of this paper. We list some notable features of $F_*$ below:

(i) For a given target state,  $F_*$ gives a fundamental limit on the state preparation under stochastic control errors.
Namely, $F_*$ clarifies the set that the controlled quantum system can or cannot reach, under given control errors and control time.

(ii) $F_*$ can be calculated without needing the full dynamics of the quantum system, 
which does not necessitate to solve any equations. 

(iii) If the assumption $\hat{B}^2(t)=\gamma^2(t)\hat{I}$ is not imposed, $F_*$ can be generalized as follows (see Appendix A for the proof):
\begin{equation}
\mathbb{E}[F(T)]  \geq F'_*:={\rm exp}\left(-\int^T_0 \sum_k\| \hat{B}_k(t) \|^2_{\rm F}dt\right).
\end{equation}

However, it should be noted that the tightness of $F'_*$ is drastically weaker than 
$F_*$, e.g., when $\hat{B}$ satisfies $\hat{B}^2=\gamma^2\hat{I}$, $F_*=e^{-\gamma^2t}$ while $F'_*=e^{-N\gamma^2t}$.

\section{Examples}

\subsection{One-qubit}

We here examine how tight the obtained bound  $F_*$ is in practice through a simple example. 
We consider the one-qubit system such as a two level atom consisting of 
the excited state $\ket{0}=[1, 0]^\top$ and the ground state $\ket{1}=[0, 1]^\top$.
We aim to drive the state from $\ket{\psi_0}=\ket{1}=[0, 1]^\top$ to 
$\ket{\psi(T)}=\ket{0}=[1, 0]^\top$ under the following setting:
\begin{equation}
\hat{H}=u(t) \hat{S}_y, \ \ \hat{B}=\gamma(t) \hat{S}_x,
\end{equation}
where
$\hat{S}_x=\ket{0}\bra{1}+\ket{1}\bra{0}$, $\hat{S}_y=i(\ket{1}\bra{0}-\ket{0}\bra{1})$ 
and $\hat{S}_z=\ket{0}\bra{0}-\ket{1}\bra{1}$ are the Pauli matrices.
$\hat{H}$ and $\hat{B}$ represent the control Hamiltonian and the control error, respectively.
$\hat{B}$ fluctuates the state vector around the $x$-axis with frequency $\gamma(t)$.
For simplicity, we assume that $u(t)$ and $\gamma(t)$ are time-independent $u(t)=u$ and $\gamma(t)=\gamma$.
Then, the lower bound $F_*$ is calculated as $F_*={\rm exp}\{-\gamma^2\pi/(2u)\}$.
Also, by solving the equation 
$d\ket{\psi(t)}/dt=-iu\hat{S}_y\ket{\psi(t)}$, we obtain the exact control time $T=\pi/(2u)$.
Moreover, assuming $2u>\gamma^2$, the average fidelity $\mathbb{E}[F(T)]$ can be analytically calculated as follows:

\begin{equation}
\mathbb{E}[F(T)]=\frac{1}{2} \left\{ 1+ \frac{e^{-\gamma^2T}}{\sqrt{D }}
\left( \gamma^2\sin(\sqrt{ D}T)-\sqrt{D}\cos(\sqrt{D }T) \right) \right\}, 
\end{equation}
where $D=4u^2-\gamma^4$ (see Appendix. B for the proof).

Figure 1(a) depicts $F_*$ and  $\mathbb{E}[ F(T) ]$ as a function of $\gamma$, in unit of $u=1$.
We find that $F_*$ is effective for  small $\gamma$.
Further, let us focus on the ratio of $\mathbb{E}[F(T)]$ and $F_*$.
When $\gamma$ is fixed, as the energy spent for the control increases (i.e., the control time becomes faster), 
$F_*$ becomes tighter  [Fig. 1 (b)]. 
From this result, we can say that a rapid state control for the quantum system is efficient to make the influence of the control errors small.

Next, we consider a time-dependent rotation error modeled by 
\begin{equation}
\hat{B}(t)=\gamma \left(\cos(\omega t)\hat{S}_x +\sin(\omega t)\hat{S}_y \right),
\end{equation}
which are common in the current quantum computing experiments.
$\omega$ is the positive constant representing the rotation frequency.
Under the same Hamiltonian and the initial and final states, 
$F_*$ has the same expression as above and gives a slightly tighter bound for all $\gamma$, 
compared than the case of $\hat{B}=\hat{S}_x$, 
as illustrated in Fig.1 (a).	
Thus, it can be seen that the lower bound $F_*$ works well for such realistic time-dependent control errors.

\begin{figure}[tb]
\centering
\includegraphics[width=12cm]{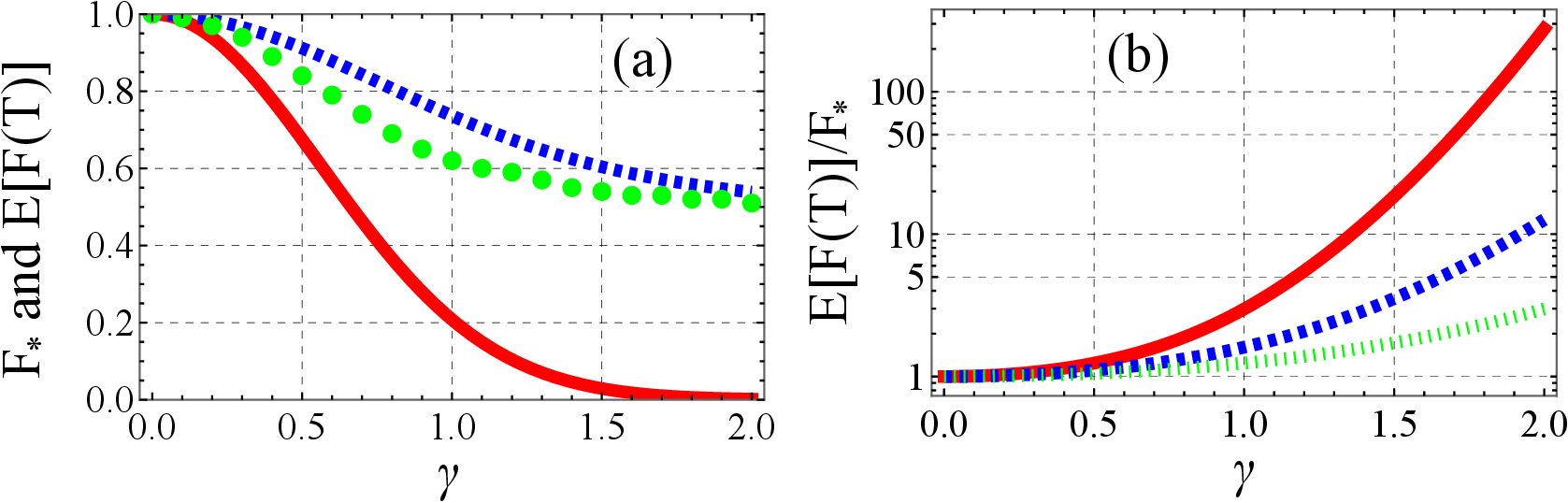}
\caption{ (a) Plots of the lower bound $F_*$ (solid red line) and $\mathbb{E}[ F(T) ]$
for simple error $\hat{B}=\gamma\hat{S}_x$ (blue dashed line) and rotation error 
$\hat{B}(t)=\gamma \left(\cos(\omega t)\hat{S}_x +\sin(\omega t)\hat{S}_y \right)$ (green dots), as a function of $\gamma$ when $u=1$ and $T=\pi/(2u)$. 
(b) Ratio $\mathbb{E}[F(T)]/F_*$ as a function of $\gamma$, for $u=1$ (red solid line), $2$ (blue dashed line), and $4$ (green dotted line).
}
\end{figure}

\subsection{Two-qubit}

Next, we study the two-qubit system under stochastic noise.
Let us focus on the SWAP operation exchanging the states of qubits 1 and 2, 
which is of particular use in the scenario of quantum information such as quantum Fourier transform (QFT) \cite{Nielsen}.
It is realized by the SWAP gate 

\begin{eqnarray}
 \hat{U}_{\rm SWAP}  =
 \left[  \begin{array}{cccc}
 1 & 0 & 0 & 0   \\ 
0 & 0 & 1 & 0   \\
0& 1 & 0 &0   \\
0 & 0 & 0& 1\end{array}
\right],  
\end{eqnarray}

and  $U_{\rm SWAP}$ is generated by the time-independent Hamiltonian
\begin{equation}
\hat{H} =\frac{u}{2}\left( \hat{S}_x\otimes \hat{S}_x + \hat{S}_y\otimes \hat{S}_y + \hat{S}_z\otimes \hat{S}_z \right).
\end{equation}

We assume that the initial state is $\ket{\psi_0}=\ket{+}\otimes \ket{0}$ where $\ket{+}=(\ket{0}+\ket{1})/\sqrt{2}$, and the final state is given by 
\begin{equation}
\ket{\psi(T)}=U_{\rm SWAP}\ket{\psi_0}= \ket{0}\otimes \ket{+}.
\end{equation}
It takes the time $T=\pi/(2u)$ for this transformation.

Moreover, we now introduce the two types of decay processes:
\begin{eqnarray}
&&\hat{B}_g =\gamma \hat{S}^1_x\otimes \hat{S}^2_x,  \\ 
&&\hat{B}^1_{l}=\gamma \hat{S}^1_x\otimes \hat{I}, \ \ \hat{B}^2_{l}=\gamma \hat{I} \otimes \hat{S}^2_x, 
\end{eqnarray}
$\hat{B}_g$ is a global noise acting on the both qubits spontaneously.
On the other hand, $\hat{B}^1_{l}$  and $\hat{B}^2_{l}$ are local noises acting on the each qubits.
For each noise cases, the lower bound is calculated as follows:

\begin{eqnarray}
F_* = \left\{
\begin{array}{ll}
e^{-\frac{\pi\gamma^2}{2u} }  & ({\rm global\  noise})   \\
e^{-\pi\gamma^2 }  &  ({\rm local\ noise})
\end{array}
\right.
\end{eqnarray} 

As illustrated in Fig. 2(a) and (b), we find that $F_*$ works as a more effective bound for the local noise than the global one,
in particular when $\gamma$ is large. 
And also, when the noise is small, $F_*$ gives a sharper estimate compared to the one-qubit case.

\begin{figure}[tb]
\includegraphics[width=12cm]{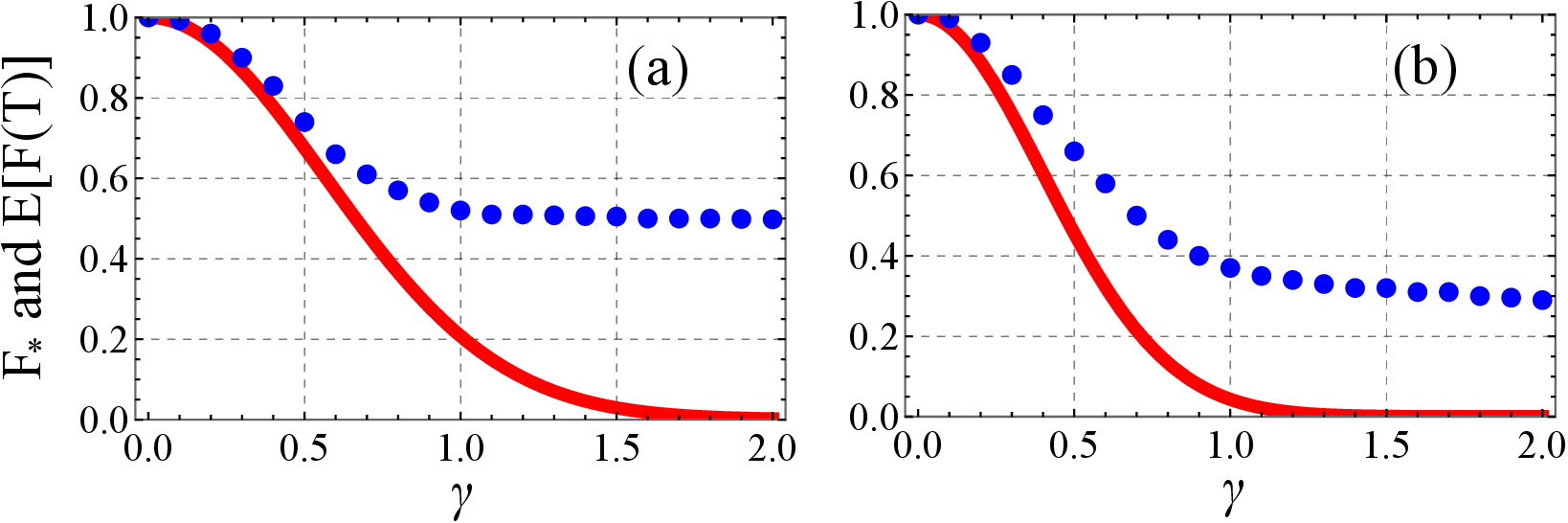}
\caption{  Plots of the lower bound $F_*$ (solid red line) and the simulated values  of $\mathbb{E}[ F(T) ]$ (blue dots) 
for (a) global noise $\hat{B}_{g}$ and (b) local one $(\hat{B}^1_{l}, \hat{B}^2_{l})$ when $u=1$ is fixed.}
\end{figure}

\subsection{Atomic ensemble}
As a further example, we study an atomic ensemble system composed of $N$ identical qubits.
We consider a case where the system is controlled by any Hamiltonian under a stochastic error.
We specify the general form representing the global noise and local one as follows:

 \begin{eqnarray}
&&\hat{B}_{g}=\gamma(t)\hat{S}^1_{x}\otimes \hat{S}^2_{x} \otimes \cdots \otimes \hat{S}^N_{x},  \\
&&\hat{B}^1_{l}=\gamma(t)\hat{S}^1_{x} \otimes \hat{I}\otimes \cdots \otimes \hat{I}, \ \cdots, \ 
\hat{B}^N_{l}=\gamma(t)\hat{I} \otimes \cdots \otimes \hat{S}^N_{x},
\end{eqnarray}
For each noises, $F_*$ is calculated as follows:
\begin{eqnarray}
F_* = \left\{
\begin{array}{ll}
e^{-\int^T_0 \gamma^2(t)dt } & ({\rm global \ noise})   \\
e^{-N \int^T_0 \gamma^2(t)dt } & ({\rm local \  noise})
\end{array}
\right.
\end{eqnarray}

From this results, $F_*$ is more effective for the global noise than the  local noise, and
also we may say that the local noise acts the system more seriously than the global noise.
Therefore, if the system is subjected to the local noise, it is required for highly suppressing the magnitude of the noise.

\section{Estimation for obtaining the target state}

Here we quantitatively investigate how likely it is to obtain the target state, in the presence of stochastic noise.
Let us expand the two final states $\ket{\psi(T)}$ and $\ket{\phi(T)}$ as follows:
 \begin{eqnarray}
\ket{\psi(T)} &=& \sum_{i=1}^n p_i \ket{i},   \\
\ket{\phi(T)} &=&  \sum_{i=1}^n q_i \ket{i},
\end{eqnarray} 
where $\left\{ \ket{n}, n=1,\cdots, m, \cdots, n \right\}$ is the measurement basis.
In the following, the derivation is based on projective measurement.
We assume that the target state is the $m$th eigenstate $\ket{m}$ of $\ket{\psi(T)}$ in the measurement basis and its probability amplitude is given by
 \begin{equation}
|p_m|^2=1-\epsilon^2,
\end{equation} 
where $0\leq \epsilon < 1$ is the acceptable value.
Let us calculate the upper bound of the overlap $|\langle \psi(T)|\phi(T)\rangle|$ as follows:

\begin{eqnarray}
|\langle \psi(T)|\phi(T)\rangle| &=& \left| \left(\sum^n_{i=1} p^*_i\bra{i} \right)  \left(\sum^n_{i=1}q_i\ket{i} \right)  \right| 
\leq  \sum^n_{i=1}|p_iq_i| \nonumber  \\
&\leq& \sqrt{1-\epsilon^2}|q_m|+\sum^n_{(i \neq m)}|p_iq_i|  \nonumber  \\
&\leq& \sqrt{1-\epsilon^2}|q_m| +\epsilon\sqrt{1-|q_m|^2}  
\end{eqnarray} 
where the Cauchy-Schwarz inequality was used. 
Therefore, combining Eq. (\ref{theorem}), the probability amplitude of the $m$th eigenstate under stochastic noises has the following lower bound:
 \begin{equation}
 \label{qbound}
|q_m| \geq q_*:= \sqrt{1-\epsilon^2}  e^{-\frac{1}{2}\int^T_0\sum_k\gamma^2_k(t)dt}
 + \epsilon\sqrt{1-e^{ -\int^T_0\sum_k\gamma^2_k(t)dt} }.
\end{equation} 

$q_*$ is always greater than or equal to zero $q_*\geq 0$, and thus, $q_*$ gives a meaningful information on state preparation. 
If we set $\epsilon=0$, we recover the result (\ref{theorem}) as $|q_m|^2 \geq \exp\left( -\int^T_0\sum_k\gamma^2_k(t)dt\right)$.
When $\epsilon$ is fixed for a certain value, $q_*\to\epsilon$ in the limit $\int^T_0\sum_k\gamma^2_k(t)dt\to \infty$; in this case, if $\epsilon=0$, it is impossible to completely obtain the desired target state.

\section{Conclusion}
In this paper, we have presented the lower bound of the fidelity between the two 
controlled quantum systems in the presence and absence stochastic control errors.
The derived bound can be straightforwardly calculated and used to derive the theoretical estimation for obtaining the target state.
We would like to emphasize that this bound clarifies a set that the quantum system can or cannot reach.
That is to say, once the stochastic noise $\hat{B}(t)$ that satisfies the condition $\hat{B}^2(t)=\gamma^2(t)\hat{I}$ is specified,
we can immediately grasp how much the quantum state can approach the target at least.
Also, we have presented the another lower bound $F'_*$ without imposing this condition, 
but its effectiveness is much weaker than $F_*$.
Thus, the derivation of a general and useful bound without limiting the type of noises is of our  interest. 
Finally, aside from these results, there is the problem that there is no methodology for dealing with stochastic errors.
An important remaining work is to develop a method to overcome stochastic control errors.

This work was supported by MEXT Quantum Leap Flagship Program Grant 
Number JPMXS0118067285 and JPMXS0120319794. 
Also it is a pleasure to thank Manaka Okuyama for insightful discussion.

\appendix

\section{Derivation of the lower bound for general noise}

We derive the generalized lower bound $F_*'$ without imposing the assumption $\hat{B}^2_k(t)=\gamma^2_k(t)\hat{I}$.
First, we present a time evolution of the operator representing the pure state $\hat{\rho}(t)=\ket{\phi(t)}\bra{\phi(t)}$:

\begin{eqnarray}
\label{drho}
d \hat{\rho}(t) &=& d\ket{\phi(t)}\cdot \bra{\phi(t)} +\ket{\phi(t)}d\bra{\phi(t)}+d\ket{\phi(t)}\cdot d\bra{\phi(t)}   \nonumber  \\
&=& \left(-i\hat{H} (t)dt-\frac{1}{2}\sum_k \hat{B}^2_k(t)dt -i\sum_k\hat{B}_k(t)\bullet dW_k(t) \right) \ket{\phi(t)}\bra{\phi(t)}   \nonumber  \\
&&\ \ \ +\ket{\phi(t)}\bra{\phi(t)} \left(i \hat{H}(t)dt-\frac{1}{2}\sum_k\hat{B}^2_k(t)dt +i\sum_k\hat{B}_k(t)\bullet dW_k(t) \right)   \nonumber  \\
&&\ \ \ +\left( -i\hat{H}(t)dt-\frac{1}{2}\sum_k\hat{B}^2_k(t)dt -i \sum_k\hat{B}_k(t)\bullet dW_k(t) \right)\ket{\phi(t)}\bra{\phi(t)} \nonumber \\
&&\ \ \ \times \left(i\hat{H}(t)dt-\frac{1}{2}\sum_k\hat{B}^2_k(t)dt   +i \sum_k \hat{B}_k(t)\bullet dW_k(t)  \right)  \nonumber  \\
&=& -i[\hat{H}(t), \hat{\rho}(t)]dt +\sum_k\mathcal{D}[\hat{B}_k(t)]\rho(t)dt -i\sum_k[\hat{B}_k(t), \hat{\rho}(t)]\bullet dW_k(t),
\end{eqnarray}
where we used $dt^2=dtdW(t)=dW(t)dt=0$ and defined $\mathcal{D}[\hat{B}]=\hat{B}\hat{\rho} \hat{B}-\hat{B}^2\hat{\rho}/2-\hat{\rho} \hat{B}^2/2$.
In the same manner, the time evolution of $\hat{\sigma}(t)=\ket{\psi(t)}\bra{\psi(t)}$ is given by

\begin{equation}
\label{dmu}
\frac{d \hat{\sigma}(t) }{dt}=-i[\hat{H}(t), \hat{\sigma}(t)].
\end{equation}

Using (\ref{drho}) and (\ref{dmu}), we calculate the infinitesimal change of the fidelity $F(t)={\rm Tr}[\hat{\rho}(t) \hat{\sigma}(t)]$ as follows:

 \begin{eqnarray}
dF(t) &=& {\rm Tr}[d \hat{\rho}(t)\hat{\sigma}(t) + \hat{\rho}(t)d \hat{\sigma}(t)]     \nonumber  \\
&=& {\rm Tr}\left\{ \left( -i[\hat{H}(t), \hat{\rho}(t)] +\sum_k\mathcal{D}[\hat{B}_k(t)] \hat{\rho}(t) \right)\hat{\sigma}(t) \right\}dt 
 + {\rm Tr}\left\{ \hat{\rho}(t) \left( -i[ \hat{H}(t), \hat{\sigma}(t)] \right) \right\} dt  \nonumber \\
&&\ \ \  -i\sum_k {\rm Tr} \left( [ \hat{B}_k(t), \hat{\rho}(t)] \right) \bullet dW_k(t)    \nonumber  \\
&=& -\sum_k{\rm Tr}\left[\hat{ B}^2_k(t)\hat{\rho}(t) \hat{\sigma}(t)\right]dt 
+\sum_k{\rm Tr}\left[ \hat{B}_k(t) \hat{\rho}(t)\hat{B}_k(t)\hat{\sigma}(t) \right]dt   \nonumber  \\
&& \ \ \ -i\sum_k {\rm Tr}\left( [\hat{B}_k(t), \hat{\rho}(t)] \right) \bullet dW_k(t)    \nonumber \\
&\geq& -\sum_k\| \hat{B}_k(t) \|^2_{\rm F}  F(t)dt  -i\sum_k {\rm Tr} \left( [\hat{B}_k(t), \hat{\rho} (t)]\right) \bullet dW_k(t),
\end{eqnarray}

where we have used  the inequality 
${\rm Tr}(\hat{A}\hat{B})\leq {\rm Tr}(\hat{A}){\rm Tr}(\hat{B})$ for positive semidefinite matrices $\hat{A}$ and $\hat{B}$.
Taking the ensemble average of (A3), we thus have

\begin{equation}
\frac{d\mathbb{E}[F(t)]}{dt}  \geq -\sum_k \| \hat{B}_k(t) \|^2_{\rm F}\mathbb{E}[F(t)].
\end{equation}

Then, by integrating the above from $0$ to $T$, we obtain the result given in (\ref{theorem}):
\begin{equation}
\mathbb{E}[F(T)]  \geq F'_*={\rm exp}\left(-\int^T_0 \sum_k \| \hat{B}_k(t) \|^2_{\rm F}dt\right).
\end{equation}

\section{Detailed calculations of average fidelity}

To obtain the exact solution of the average final fidelity $\mathbb{E}[F(T)]$,  
it is needed to calculate $\mathbb{E}[\hat{\rho}(T)]=\mathbb{E}[\ket{\phi(T)}\bra{\phi(T)}]$, because 
\begin{equation}
\mathbb{E}[F(T)]=\mathbb{E}[|\langle \phi(T)|\psi(T)\rangle|^2]={\rm Tr}\{\mathbb{E}[\hat{\rho}(T)]\hat{\sigma}(T)\}.
\end{equation}

For the setup $(\hat{H}, \hat{B})=(u\hat{S}_y, \gamma \hat{S}_x)$,
 by using Eq. (\ref{drho}), the time evolution of $\mathbb{E}[\hat{\rho}(t)]$ is described as follows:
\begin{equation}
\frac{d\mathbb{E}[ \hat{\rho}(t)] }{dt}=-i[u\hat{S}_y, \mathbb{E}[ \hat{\rho}(t)]] +\mathcal{D}[\gamma \hat{S}_x]\mathbb{E}[\hat{\rho}(t)].
\end{equation}
This equation leads to the two differential equations:

\begin{eqnarray}
&&\frac{d \mathbb{E}[z(t)] }{dt} =-2u\mathbb{E}[x(t)]-2\gamma^2\mathbb{E}[z(t)] \\
&&\frac{d \mathbb{E}[x(t)] }{dt} =2u\mathbb{E}[z(t)],
\end{eqnarray}
where we defined $x(t)={\rm Tr}[\hat{\rho}(t) \hat{S}_x]$  and $z(t)={\rm Tr}[\hat{\rho}(t) \hat{S}_z]$.
In order to calculate $\mathbb{E}[F(T)]$, it is necessary to obtain the information on $\mathbb{E}[z(t)]$ only, 
because $\mathbb{E}[F(T)]=\langle 0| \mathbb{E}[\hat{\rho}(T)] |0\rangle=(1+\mathbb{E}[z(T)])/2$.
From this two equations,  we have the second-order differential equation:
\begin{equation}
\frac{d^2 \mathbb{E}[z(t)] }{dt^2} =-2\gamma^2\frac{d \mathbb{E}[z(t)] }{dt}-4u^2 \mathbb{E}[z(t)].
\end{equation}
By assuming $2u>\gamma^2$ and using the initial conditions $\mathbb{E}[z(0)]=-1$ and $\mathbb{E}[x(0)]=0$, 
the solution of this equation is thus
\begin{equation}
\mathbb{E}[z(T)] =\frac{e^{-\gamma^2t}}{\sqrt{D}}\left( \gamma^2\sin(\sqrt{D}t)-\sqrt{D}\cos(\sqrt{D}t) \right), \ \ \ D=4u^2-\gamma^4.
\end{equation}
Therefore,  we obtain the analytical solution of $\mathbb{E}[F(T)]$:
\begin{equation}
\mathbb{E}[F(T)]=\frac{1}{2}\left(1+ \frac{e^{-\gamma^2T}}{\sqrt{D}}\left( \gamma^2\sin(\sqrt{D}T)-\sqrt{D}\cos(\sqrt{D}T) \right)\right).
\end{equation}


\end{document}